\documentclass[iop]{emulateapj}
\usepackage{color}
\usepackage{natbib}
\slugcomment{}

\shorttitle{Potassium in globular cluster stars}
\shortauthors{Carretta et al.}

\begin{document}
\title{Potassium in globular cluster stars: comparing normal clusters to
the peculiar cluster NGC~2419\altaffilmark{1}}

\author{E. Carretta\altaffilmark{2},
R.G. Gratton\altaffilmark{3},
A. Bragaglia\altaffilmark{2},
V. D'Orazi\altaffilmark{4,5},
S. Lucatello\altaffilmark{3},
A. Sollima\altaffilmark{2},
C. Sneden\altaffilmark{6}
}

\altaffiltext{1}{Based on data collected at the ESO telescopes under 
programmes 085.D-0205 and 165.L-0263, and at the McDonald 2.7 m telescope.}
\altaffiltext{2}{INAF, Osservatorio Astronomico di Bologna, via Ranzani 1,
       40127,  Bologna,  Italy. eugenio.carretta@oabo.inaf.it,
       angela.bragaglia@oabo.inaf.it, antonio.sollima@oabo.inaf.it}
\altaffiltext{3}{INAF, Osservatorio Astronomico di Padova, vicolo
       dell'Osservatorio 5, 35122 Padova,  Italy. raffaele.gratton@oapd.inaf.it
       sara.lucatello@oapd.inaf.it}
\altaffiltext{4}{Department of Physics and Astronomy, Macquarie University,
Balaclava rd, North Ryde, NSW 2109, Australia, valentina.dorazi@mq.edu.au}
\altaffiltext{5}{Monash Centre for Astrophysics, School of Mathematical
Sciences, Building 28, Monash University, VIC 3800, Australia}
\altaffiltext{6}{University of Texas at Austin, Department of Astronomy, 
Austin, TX, 78712, USA, chris@verdi.as.utexas.edu}

\begin{abstract}

Two independent studies recently uncovered two distinct populations 
among giants in the distant, massive globular cluster (GC) NGC~2419.
One of these populations has normal magnesium (Mg) and potassium (K) 
abundances for halo stars: enhanced Mg and roughly solar K. 
The other population has extremely depleted Mg and very enhanced K.  
To better anchor the peculiar NGC~2419 chemical composition, we have 
investigated the behavior of K in a few red giant branch stars in 
NGC~6752, NGC~6121,  NGC~1904, and $\omega$~Cen. 
To verify that the high K abundances are intrinsic and not due to some 
atmospheric features in giants, we also derived K abundances in less evolved 
turn-off and subgiant stars of clusters 47~Tuc, NGC~6752, NGC~6397, and NGC~7099. 
We normalized the K abundance as a function of the cluster metallicity using 
21 field stars analyzed in a homogeneous manner.
For all GCs of our sample, the stars lie in the K-Mg abundance plane on the 
same locus occupied by the Mg-normal population in NGC~2419 and by field stars.
This holds both for giants and less evolved stars. 
At present, NGC~2419 seems unique among GCs. 

\end{abstract}


\section{Introduction}

Galactic globular clusters (GCs) show distinctive chemical and 
photometric signatures that can vary over a large range from cluster to cluster.
The complex chemical composition patterns of GCs are nearly non-existent 
among halo field stars of our Galaxy \citep{gra00} and in its satellites 
\citep{tol09}. 
In most cases the basic nucleosynthesis is attributable to light-element 
proton-capture reactions in H-burning at high temperature, which leads
eventually to large stellar surface variations in the abundances of
C, N, O, Na, Al, and in carbon isotopic ratios. 
These variations can be most easily understood in terms 
of multiple populations within each cluster; see recent reviews by 
\cite{gra12} and \cite{pio10} for more detailed discussion.

The most notable  chemical signature of multiple populations, the Na-O 
anti-correlation (discovered by  the Lick-Texas group, see \citealt{kra94}), 
was recently proposed as a definition of a genuine GC \citep{car10a}. 
Extensive surveys of light element abundances in GCs (see e.g. 
\citealt{car06,car10a}) can be understood with a common series of evolutionary
events:  formation of a first stellar generation, pollution of 
intra-cluster medium by a fraction of the most massive stars of this 
generation, formation of a more spatially compact second generation, with the
loss of most of the first stellar population (see e.g. \citealt{der08,decr08}).
These actions occurred in each GC in slightly different ways.

A notable example of GC abundance anomalies is NGC~2419. 
This is the third most massive cluster in our Galaxy ($M_V = -9.42$, \citealt{har96}). 
Unfortunately it is located at a very large distance 
\citep[about 88 kpc,][]{dic11}, and thus even its brightest members
are faint ($V$~$>$ 17) for high-resolution spectroscopy. 
Despite this daunting observational difficulty, the chemical compositions of 
many NGC~2419 red giants have been recently investigated by 
\cite{coh10,coh11,coh12} and \cite{muc12}.
Studies based on both medium-resolution DEIMOS \citep{muc12} and high-resolution
HIRES \citep{coh12} spectra have revealed two populations of stars in this
cluster that are characterized by large differences in magnesium and potassium
abundances. 
One population is made of giants with normal low-metallicity overabundances
of [Mg/Fe] and nearly solar [K/Fe] abundance ratios, and the other 
includes stars with large enhancements of K and huge depletions of Mg.  
Stars of the latter composition have not been observed to date in other GCs. 
The reality of this anti-correlation has been discussed thoroughly by
\cite{muc12} and \cite{coh12}. 
Possible spurious effects due to the analysis (non-LTE corrections, 
contamination from telluric lines, strong velocity fields in the upper 
stellar atmosphere, uncertainties in the adopted atmospheric parameters) 
were thoroughly scrutinized in those papers, and disregarded as possible 
sources of this unusual Mg/K abundance pattern.

The findings by Cohen et al. and Mucciarelli et al. have stimulated 
theoretical studies on this peculiar cluster.
In the framework of multiple populations in GCs, \cite{ven12} propose
that a Mg-K anti-correlation might be produced by the simultaneous activation 
of the Mg-Al-Si and Ar-K cycles in massive asymptotic giant branch (AGB) stars
at the low metallicity of NGC~2419, provided that the relevant cross-sections 
and/or the efficiency of hot bottom burning  are larger than 
commonly believed.
\cite{coh12} found a rather clear correlation between Na and Al abundances 
in NGC~2419, which likely results from the coupling of the NeNa and MgAl 
cycles that enhance Na and Al in typical GCs (e.g. \citealt{den89,lan93}, 
see also \citealt{car09} for a large survey of GCs). 
However, they conclude that the K-Mg anti-correlation should have a different 
origin, because the Na-rich/Al-rich stars (the second-generation stars formed 
in normal GCs, see \citealt{car10a}) do not correspond to the stars with 
strongly depleted Mg and enhanced K abundances.
Possibly this argument is not as conclusive as proposed by \cite{coh12}, since
\cite{ven12} show that stars that are extremely depleted in Mg 
(and K-rich) may be not very Al-rich, because at very high temperature 
Mg is almost entirely transformed into Si rather than Al.

More K abundances in GCs would be clearly welcome.
We wish to confirm, using new data available to us, the growing evidence 
that NGC~2419 is really unique among the GCs in our Galaxy and discuss if 
the observed K-Mg  anti-correlation should be included in the set of 
signatures typical of proton-capture burning in normal GCs.  
This second issue can be treated following various paths. 
First, to better understand the nucleosynthesis role of this
poorly-studied element, it is important to derive its abundance in the
same stars as those with the full set of light element (at least 
O, Na, Mg, Al) abundances.
\cite{tak09} measured the K abundance in five bright red giant
branch (RGB) stars in each of the three clusters M~4, M~13 and M~15.
Unfortunately, they did not provide abundances of Mg or other proton-capture
elements. 
Their analysis only showed a remarkable small scatter in [K/Fe] ratios
within each cluster, apart from one star in M~4 and one in M~13, whose peculiar
abundances were explained as an effect of the increased velocity field in the
upper atmospheric layers (the net result is to increase the strength of the
already saturated K resonance lines). 
Note, however, that a reanalysis by \cite{coh12} of the discrepant giant 
in M~13 found normal [Mg/Fe] and [K/Fe] values for that star. 
Recently, \cite{roe11} measured the K abundance in six giants
of M~22 (three for each sub-population in this peculiar cluster). 
The [K/Fe] ratio is higher in the Ca-rich, $s-$process rich group, but all 
six stars lie in the region populated by Mg-normal stars in NGC~2419.
Second, even if found in other GCs, it would be important to verify that the
K-Mg anti-correlation is a phenomenon not restricted to the photospheric
abundances of giant stars only. 
Finding such a pattern also in relatively unevolved cluster 
stars would be a guarantee that no spurious (such as chromospheric activity) 
or evolutionary (like mixing) effects are at work.

To shed light on these issues we have looked for available observational
material apt to derive accurate K abundances in (i) RGB stars, (ii) 
main sequence turn-off and subgiant stars of normal GCs, and 
(iii) field stars, both dwarfs and giants. 
We present K abundances in 25 red giants of four GCs (NGC~6752, NGC~6121, 
NGC~1904, and the most massive GC in the Galaxy, $\omega$ Cen), and in 
17 turn-off and 35 subgiants in four GCs (NGC~6752, 47~Tuc, NGC~7099, 
and NGC~6397). 
With this wealth of data we increase the evidence that NGC~2419
is  presently an unique object among GCs,  and likely represents a case where
proton-capture reactions occurred under very peculiar conditions, as advocated
by \cite{ven12}. 
Moreover, we discuss the evidence that NGC~2419 is also
different from the typical metal-poor population of dwarf galaxies.

\section{Available datasets, observations and analysis}

All K abundances discussed in the present paper are based on 
equivalent widths (EWs) of the resonance K {\sc i} line at
7698.98~\AA; the other observable resonance doublet transition at 7664.91~\AA\
is heavily affected by strong absorption telluric lines.
The relevant data are summarized in Table~1.

Spectra for RGB stars are mainly from our program (ESO 085.D-0.205) 
devoted to study Al abundances for a large sample of stars in selected GCs (see
\citealt{car12}). 
We acquired fiber-fed UVES Red Arm high-resolution ($R\sim 43,000$) spectra 
centered at 860 nm and covering the spectral range approximatively from 6730 
to 10150~\AA.  
From these spectra we only derive the K abundances for stars in NGC~6752, 
M~4, and NGC~1904; the abundances of the other elements discussed 
in the present paper for these giants were taken from \cite{car09} 
and \cite{car10c}.

Spectra for six giants in $\omega$ Cen and for all the less-evolved 
stars are from the ESO Large Program 165.L-0263 (PI Gratton). 
Description of the observations can be found in \cite{gra01,pan02,car04,car05}. 
The spectra for M~30 were acquired with the same setup 
of the other clusters. 

Spectra were reduced by the ESO personnel with the dedicated
pipelines, by extracting one-dimensional, wavelength-calibrated
spectra that were sky subtracted and shifted to zero radial velocity. 

Abundances of K were also obtained for 21 field stars with 
metallicities $-2<$[Fe/H]$<-1$~dex selected from the sample of 
stars with good parallaxes used by \cite{gra00} to study the mixing episodes 
in low mass Pop II stars. 
For that project, high S/N ($>$ 100), high resolution ($R>50,000$) 
spectra with very broad spectral coverage were obtained with the 
McDonald 2.7m telescope and Tull coud\'e echelle spectrograph 
(see \citealt{gra00} for further details).

Significant corrections for departures from the LTE assumption should be 
considered when deriving K abundances \citep[e.g.,][]{tak02,tak09}. 
The corrections are a function of line strength, because strong lines form 
at shallower optical depths, where the non-LTE effects are larger. 
They increase with decreasing surface gravity (less frequent collisions) and 
increasing temperature (larger ionizing flux). 
Since our stars span a large range of parameters, appropriate
corrections should be  considered for each case. 
We adopted non-LTE abundance corrections which are function of 
temperature, gravity, metal abundance, and EW of the K {\sc i} line from 
a multivariate interpolation through the about 900 models provided by
\cite{tak02}\footnote{$http://optik2.mtk.nao.ac.jp/\sim takeda/potassium\_nonlte$}.
The same corrections were applied to the tabulated abundances of 
\cite{coh12} and \cite{muc12}. 
For the latter, we first corrected upward by 0.3 dex their 
tabulated values of [K/Fe] to recover the LTE values.

\section{Results and discussion}

\subsection{Setting the stage: the normalization with field stars}

The derived abundances for K in our program GCs are listed in Table~1, 
while in Table~2 we list the derived abundances for field stars.
References for the adopted atmospheric parameters and abundances of the other
elements are  also listed. For each cluster, star-to-star errors were derived 
using the sensitivities of abundances to changes in the atmospheric parameters
and the internal uncertainties in each parameter as listed in the original
papers. 
For the giants in $\omega$~Cen we determined the sensitivities using the
line list adopted in all our recent analyses (from \citealt{gra03}), since they
were not provided in \cite{pan02}. 
The [K/Fe] ratios in Tables~1 and 2 include the non-LTE corrections.

\begin{figure} 
\includegraphics[width=12cm]{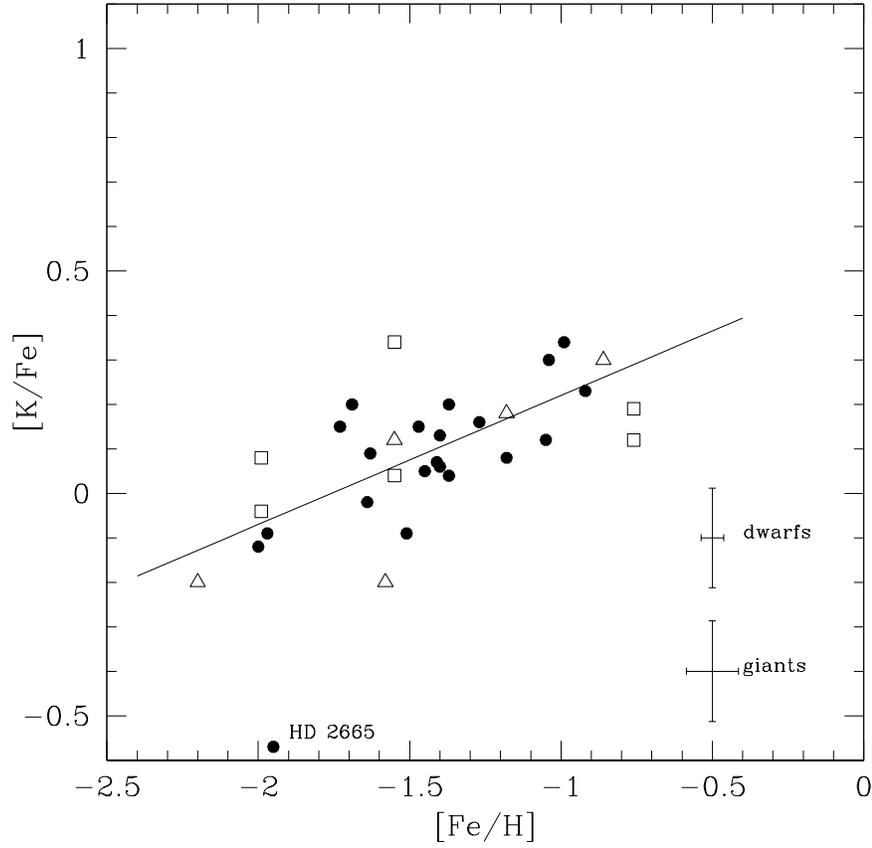}
\caption{[K/Fe] ratios as a function of the metallicity in our program stars and
in the K-poor stars of NGC~2419 (from \citealt{coh12}). Empty symbols
indicate average values for GCs (triangles: giants, squares: unevolved stars).
Filled circles: field stars, to which the error bars refer. The solid line is
the linear regression fit to the field stars, excluding star  HD 2665.}
\label{f:fig1}
\end{figure}

In Fig.~\ref{f:fig1} we show the run of the [K/Fe] ratio as a function of
the metallicity for our program stars and the K-poor group in NGC~2419 from
\cite{coh12}. 
The average value of [K/Fe] seems to be different from cluster to cluster, 
with a trend to have lower values in more metal-poor clusters. 
This trend is similar to what we found in field stars of our
sample, regardless from the evolutionary state. 
The clear trend of our [K/Fe] with [Fe/H] persists even when 
the value for HD~2665 (one of the most metal-poor stars of our sample, 
with very low [K/Fe]) is excluded.
The linear correlation coefficient is $r=0.70$\ over 21 stars, which is 
highly significant. 
Once HD~2665 is dropped, the best fit linear regression line is ${\rm
[K/Fe]}=(0.29\pm 0.07){\rm [Fe/H]} + (0.51\pm 0.09)$. 
Note that while our mean value of [K/Fe]=0.10 (r.m.s.=0.12 dex) 
agrees fairly well with literature values from \cite{tak02}, \cite{zha06}, 
and \cite{and10}, none of these studies seems to support a trend of 
[K/Fe] with [Fe/H].
To check if our trend is an artifact of our analysis, we examined stars 
in common with those previous works. 
We have no stars in common with the samples of \cite{zha06} and
\cite{and10}, and only three stars in common with the sample by 
\cite{tak02}, listed in Table~3. 
Our measured $EW$s agree well with the values of \cite{tak02}. 
Moreover, using their Table~1, we verified that all the differences 
in the [K/Fe] ratio (even those as large as in the case of stars
HD~103095 and HD~122956) can be explained by the differences in the atmospheric
parameters, the NLTE corrections and, more importantly for dwarf stars, the
treatment of damping. 
We defer a full discussion of K abundances in metal-poor field stars
to a future study, where additional effects like the impact of 3-d model atmospheres will be treated in more detail.
However, since they were obtained in homogeneous way, we may use present results as a reference against which K abundances in GCs are compared. 
Using the average [Fe/H] value of each cluster we derived an offset in 
[K/Fe] with respect to the linear fit of field stars. 
This offset was then applied to each individual value in the GC.
Note that this is only a second order effect, aimed to obtain more uniform
abundances of K, and does not affect in any significant way our main
conclusions.

\subsection{Results for globular clusters}

\begin{figure} 
\includegraphics[width=12cm]{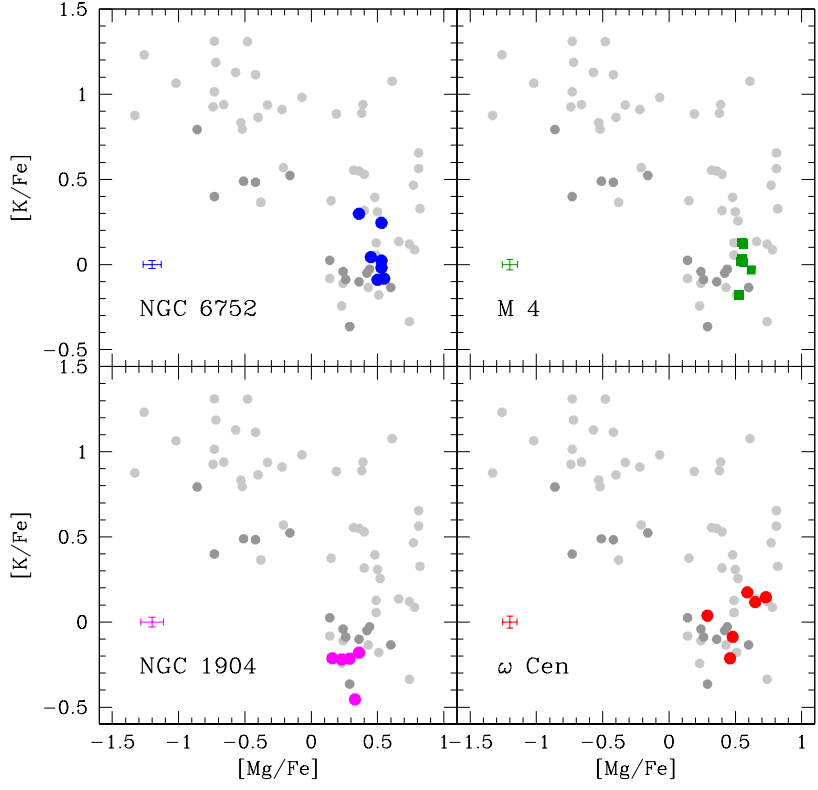}
\caption{K-Mg anti-correlation in giants of NGC~2419, from
\cite{coh12} (dark grey points) and \cite{muc12} (light grey), and in our
program cluster: RGB stars in NGC~6752 (upper left panel), M~4 (upper right),
NGC~1904 (lower left), and $\omega$ Cen (lower right) are 
superimposed on the distribution observed in NGC~2419.} 
\label{f:fig2}
\end{figure}

\begin{figure} 
\includegraphics[width=12cm]{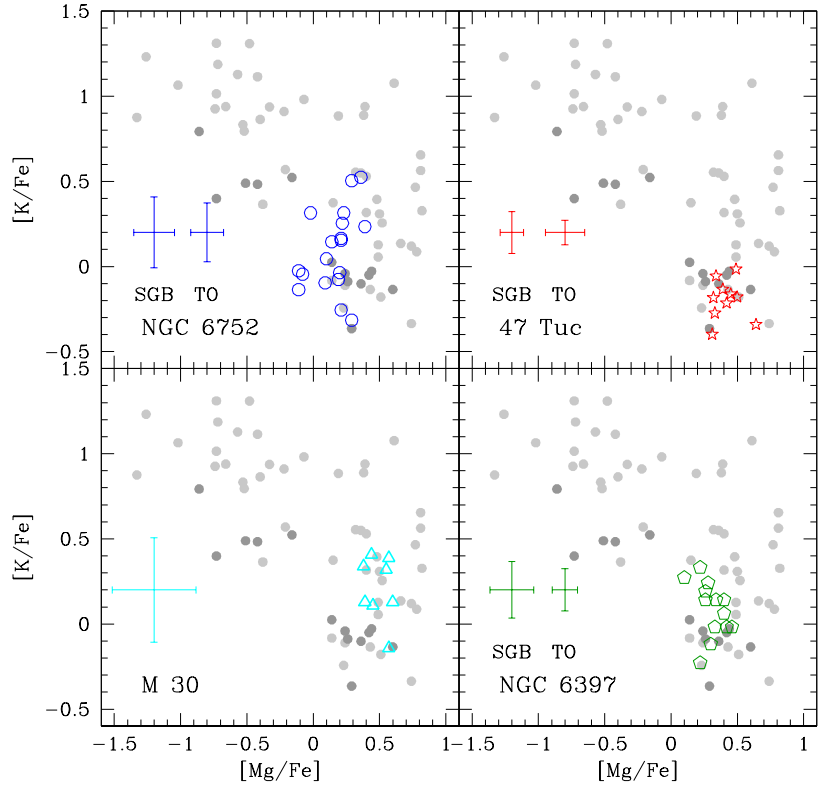}
\caption{As in Fig. 1, but this time for less evolved stars 
in our program clusters. Stars in NGC~6752 (upper left panel), 
47~Tuc (upper right), M~30 (lower left), and NGC~6397
(lower right) are shown superimposed on RGB stars in NGC~2419.}
\label{f:fig3}
\end{figure}

We focused on the star-to-star K abundance variations. 
The [K/Fe] ratios are plotted against [Mg/Fe] ratios 
in Fig.\ref{f:fig2} and Fig.~\ref{f:fig3}.
In Fig.~\ref{f:fig2} we show the abundances of giants in four clusters
(NGC~6752, M~4, NGC~1904, and $\omega$~Cen) superimposed on these
abundances for NGC~2419 from \cite{coh12} and  \cite{muc12}.
The K and Mg abundances derived for less evolved stars in NGC~6752, 
47~Tuc, M~30 and NGC~6397 are in Fig.~\ref{f:fig3}.
Our basic result is that {\it in all other GCs, stars occupy only the region
where a population with canonical Mg overabundance and moderate K abundances 
lie in NGC~2419}.
None out of 77 stars in seven different GCs share the [Mg/Fe] and [K/Fe] ratios
of the super Mg-poor/K-rich group observed in NGC~2419, not even those in
$\omega$~Cen. 
This was expected for Mg (see Fig. 10 in  \citealt{muc12}), but our study
extends this knowledge also to K abundances\footnote{A scrutiny of
unpublished K abundances was done in \cite{coh11} (see their summary), who found
no star similar to star S1131 in NGC~2419 in any other cluster. However, no
further details are given.}. 
Admittedly, our data show a hint of K-Mg anti-correlation in stars of NGC~6752; 
however the range of K abundances is very small and might be largely due to 
the possible offsets of the result we obtained for SGB stars with 
respect to TO and RGB stars. 
No trend is discernible for the remaining clusters (see e.g. the case of M~4, 
where our data confirm earlier analysis by \citealt{tak09}), whatever
evolutionary  phase is considered.

\begin{figure} 
\includegraphics[width=12cm]{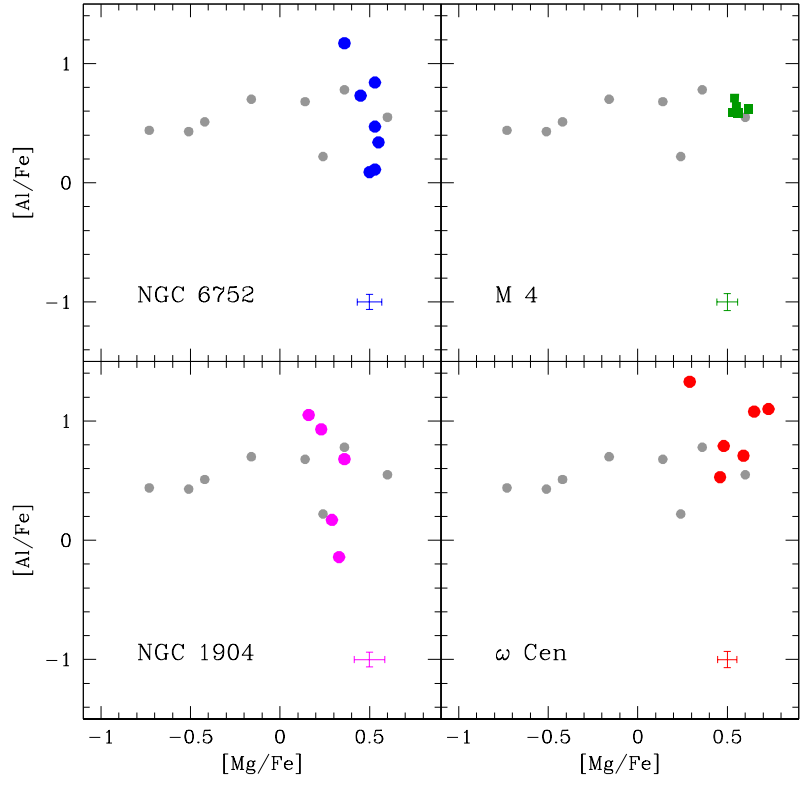}
\caption{Al-Mg in giants stars of our program clusters, superimposed on
those of NGC~2419 from \cite{coh12}. }
\label{f:fig4}
\end{figure}

\begin{figure} 
\includegraphics[width=12cm]{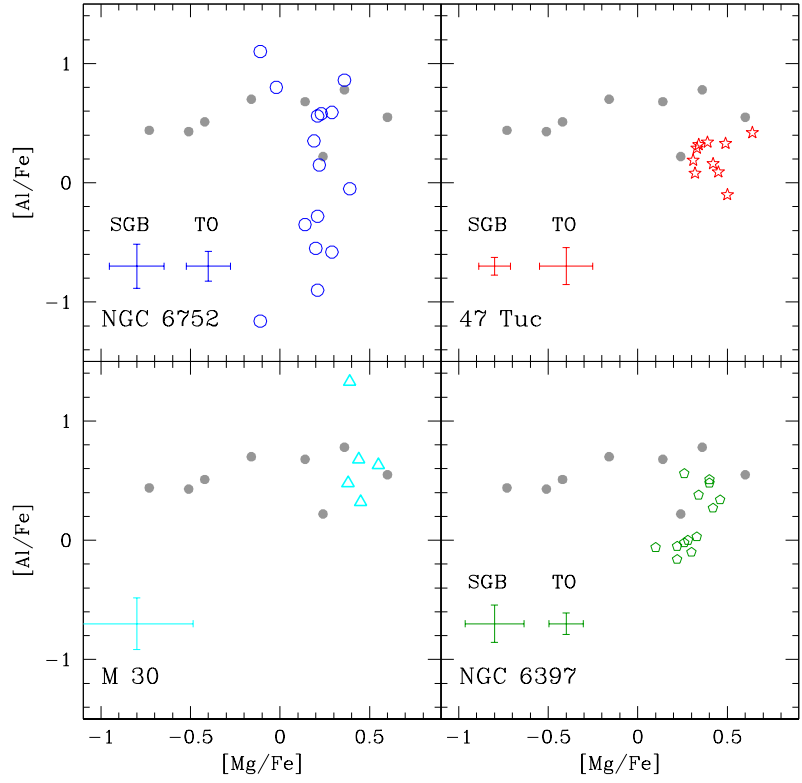}
\caption{As in Fig.3, but for unevolved stars in our program clusters.}
\label{f:fig5}
\end{figure}

In the scenario proposed by \cite{ven12}, the Mg-K anti-correlation could be 
due to the simultaneous activation of the Mg-Al-Si and Ar-K cycles, so we might 
expect a corresponding large production of Al, although further $p-$capture on
$^{27}$Al  might transform it into $^{28}$Si (see \citealt{ven12}). 
A comparison between the  abundances of the elements involved in the two cycles 
is fundamental.
The classical Al-Mg anti-correlation in our GCs is compared to the pattern in
NGC~2419 in Fig.~\ref{f:fig4} and Fig.~\ref{f:fig5}. 
Al-Mg anti-correlations of moderate extent are seen in normal clusters, 
most clearly in the more metal-poor GCs like NGC~6752 and NGC~1904. 
M~4 shows the well known small star-to-star variations in Al 
abundances \citep{iva99,mar08,car09}. 
The present sample in $\omega$~Cen does not present such anti-correlation, 
but it is not representative of the whole cluster, because half of it was 
chosen by \cite{pan02} to be on the metal-rich ``RGB-a'' branch. 
More extensive data shows that a clear Mg-Al anti-correlation is present in 
$\omega$ Cen (see e.g., \citealt{ndc95a}).  
For NGC~2419, \cite{muc12} did not measure Al abundances while 
\cite{coh12} derived Al abundances for a subset of eight giants
out of 13 in their sample. 
Al seems to be roughly constant,  or even correlated
to Mg, over the large range of Mg abundances in NGC~2419.

\section{Is the chemical inventory in NGC~2419 really unique?}

Our present results, as well as previous extensive studies 
of NGC~2419, obviously point to the fundamental question of why this peculiar 
chemical inventory is only found in this particular cluster.
More massive clusters, like $\omega$~Cen, or GCs as metal-poor as 
NGC~2419 (like NGC~6397 or M~30), do not show the same extreme
chemical pattern.

A possibility to be explored is that the pollution for the extreme and normal
populations originated from two different sources. 
Let us assume that the Mg-rich population in NGC~2419 was polluted by 
classical core-collapse supernovae (SNe), and suppose as a working hypothesis 
that the part of the stellar population with low Mg abundances was instead 
contaminated by the ejecta of a single very peculiar SN. 
A possible example of this class could be the so called pair instability 
SN (PISN). 
This is a very rare event \citep{ren12} theorized to end the life 
of a very massive population III star (see e.g. \citealt{heg05}).
The extreme rarity of these events (\citealt{ren12} identified 18 candidate
stars possibly contaminated by PISNe over a sample of 12,300 stars with
spectroscopy from SDSS) could explain why only a GC out of about 150 objects in
the Galaxy shows this peculiar pattern. 
These stars are so massive ($140 \leq M \leq 260 M_\odot$) that one single 
explosion could well have provided all the  $\sim 80 M_\odot$ of metals 
required for the proto-cluster from which NGC~2419 formed.

\begin{figure} 
\includegraphics{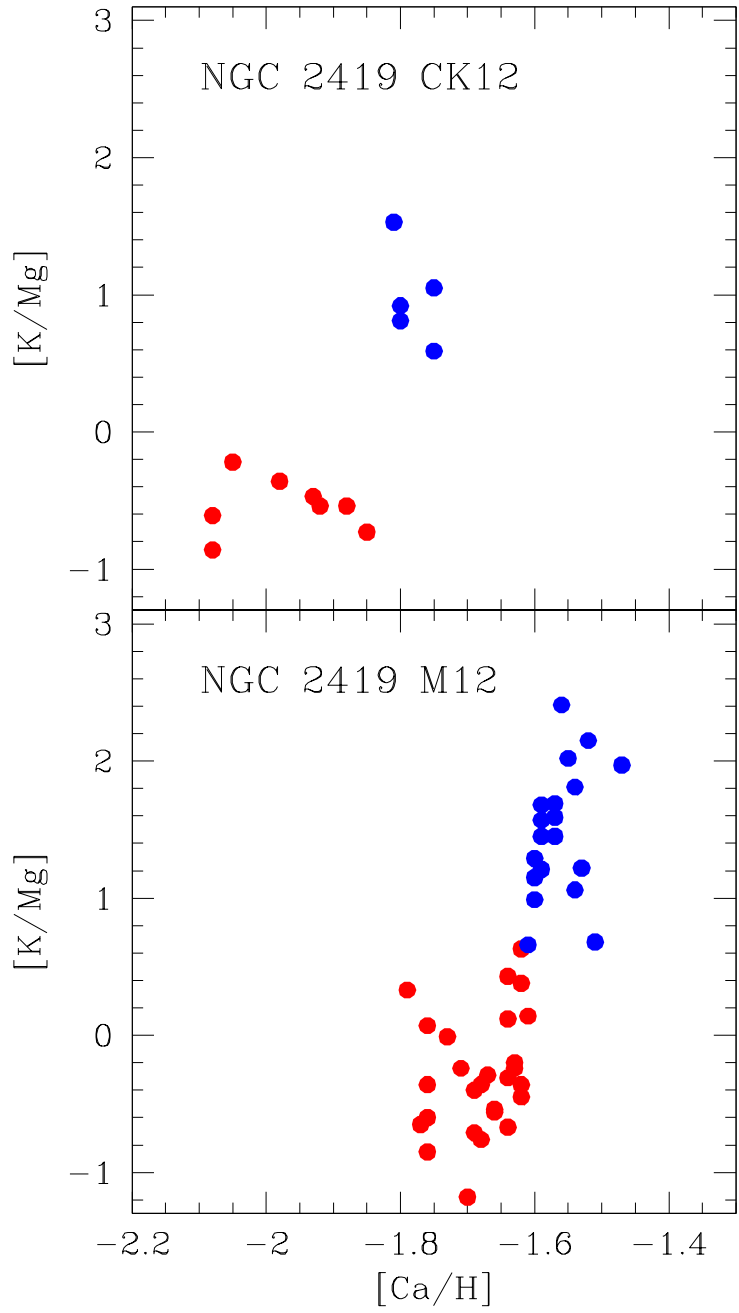}
\caption{Top panel:the [K/Mg] ratio as a function of the [Ca/H] ratio in stars
of NGC~2419 from \cite{coh12}. Bottom panel: the same, from \cite{muc12}.
Blue symbols represent stars with [Mg/Fe]$<0$.}
\label{f:fig6}
\end{figure}

Since the majority of stars incorporating a dominant contribution from PISNs are
predicted to show a strong overabundance of Ca with respect to iron, we can
check this hypothesis by looking at the run of Ca and K as a function of Mg. 
If the scenario is correct, Ca should be enhanced and Mg depleted in the putative 
population contaminated by the PISN in NGC~2419. 
This is just what we observe (see Fig.~\ref{f:fig6}) from the data of 
both \cite{coh12} and \cite{muc12}.  

However, other predictions from the nucleosynthesis associated to PISN
explosions clash with the observations for NGC~2419. 
A characteristic chemical signature associated to PISNe is a strong odd-even 
effect, which is clearly absent \citep{coh12}. 
Moreover, neutron-capture elements are
present, while they are predicted to be absent in the ejecta of PISNs.

\begin{figure} 
\includegraphics{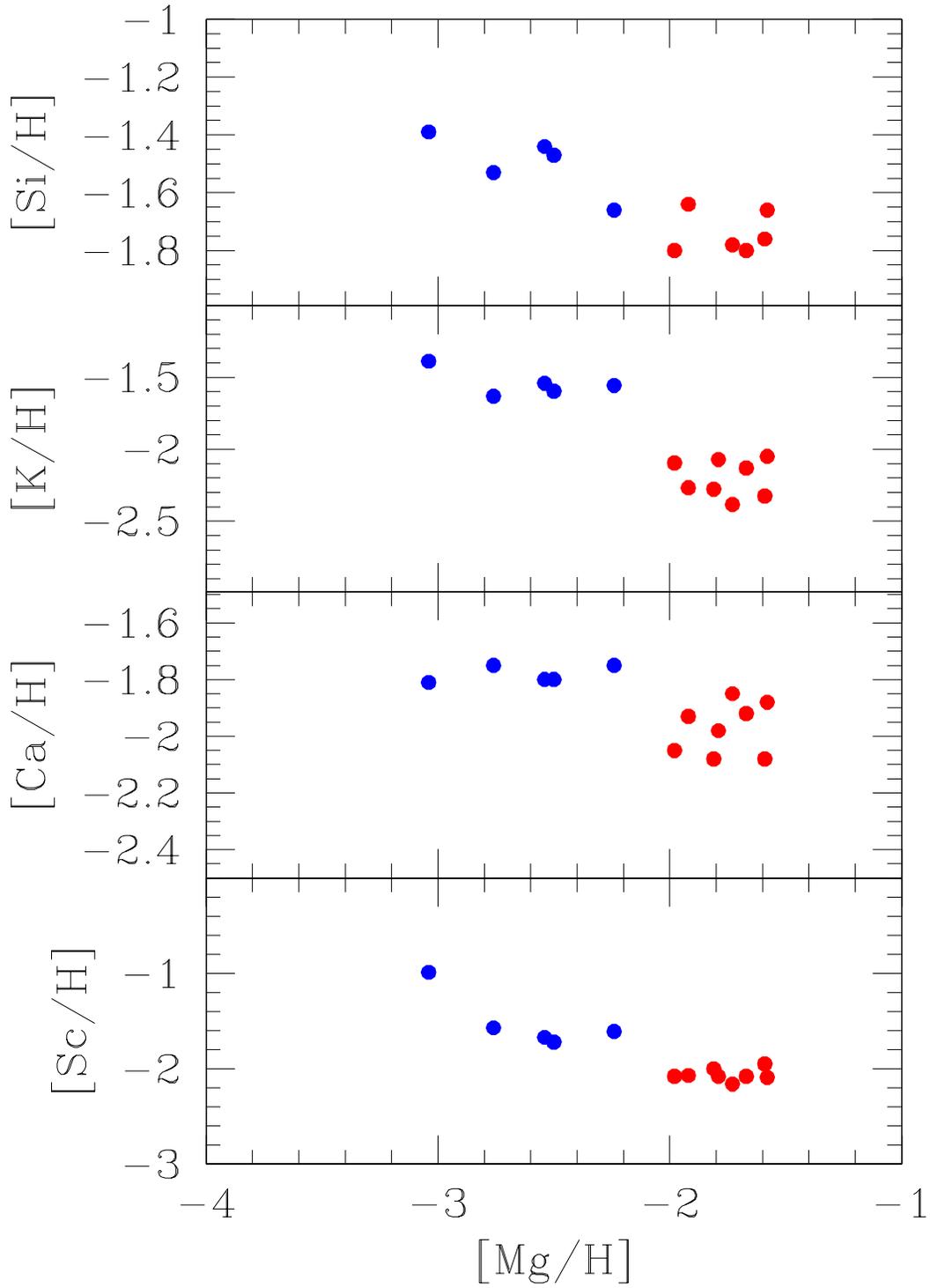}
\caption{From to to bottom: abundance ratios [Si/H], [K/H], [Ca/H], and [Sc/H] as
a function of the abundance of Mg in NGC~2419 from \cite{coh12}. Blue symbols
indicate stars with [Mg/Fe]$<0$.}
\label{f:fig7}
\end{figure}

On the other hand, the relations plotted in Fig.~\ref{f:fig6} may be explained
also by the hypothesis of proton-capture reactions occurring in a temperature
range much higher than usually observed in more normal cluster stars
(\citealt{ven12}). 
In these particular circumstances, normal intermediate steps such as
the production of Al from destruction of Mg are bypassed favoring the
synthesis of heavier elements, and the production of K, Ca, and even Sc
may be activated by p-captures on Ar nuclei. 
In this case, we should observe an anti-correlation between Mg and the 
elements enhanced in this burning, and this is just what observed in NGC~2419 
from the large dataset of species analyzed by \cite{coh12}. 
As seen in  Fig.~\ref{f:fig7}, Si, K, Ca, and Sc are nicely 
anti-correlated with Mg. 
The magnitude of the effect is larger for K and Sc, but this might simply 
be due to Ca and Si being much more abundant species than
K and Sc: transformation of even a small fraction of Ar is enough to produce
large enhancements of these last elements. 
The scenario put forward by \cite{ven12} agrees with the
observations\footnote{On the other hand, in the data of M~22 
\citep{roe11} K and Ca seem to be correlated. 
This simply corresponds to the evidence that the two populations in this 
cluster also differ in Ca, higher in the more metal-rich population (see
\citealt{mar11,roe11}). 
From the data of \cite{roe11} we also found that K is higher in the $r+s$ 
group of stars with respect to the $r$ only stars, with the
[K/Fe] ratio increasing when Sr, Y, Zr, Ba, La, Ce, Nd, and Pb (but not Eu) are
more abundant.}.

\begin{figure} 
\includegraphics[scale=0.8]{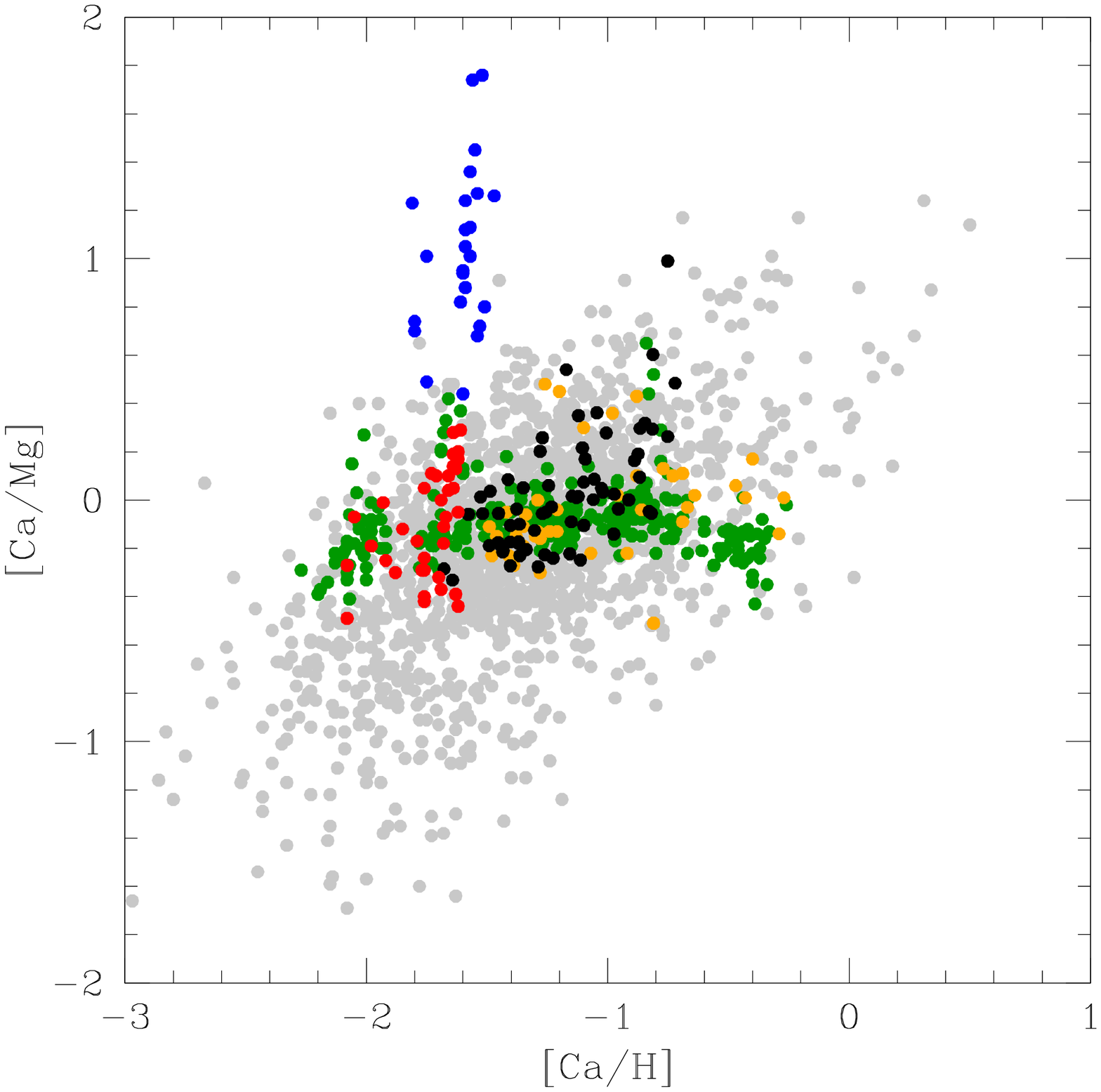}
\caption{The [Ca/Mg] ratio as a function of the Ca abundances for several
stellar populations. Grey symbols are for stars in 8 dwarf spheroidal Milky Way
satellite galaxies from \cite{kir11}; blue and red are for RGB stars in NGC~2419
with [Mg/Fe] lower and larger than 0.0 dex, respectively, from \cite{coh12} and
\cite{muc12}; orange is for giants in $\omega$ Cen from \cite{ndc95b}; black is
for giants in M~54 from \cite{car10b}, and green is for giants in 17 Galactic 
GCs from \cite{car09} (Mg) and \cite{car10c}.}
\label{f:fig8}
\end{figure}

Of course, we still need to understand why the observed extreme processing is
only observed in NGC~2419. 
This is likely not due to the small size of our sample. 
In fact, while there is still a paucity of determinations of K abundances, 
data for Ca and  Mg are available for many more stars, and can then be used 
to check whether the products of the Ar-K-Ca cycle can be seen elsewhere. 
In Fig.~\ref{f:fig8} we compare the  pattern observed in NGC~2419
(from the combined studies of \citealt{coh12} and  \citealt{muc12}) with those
of several old stellar populations.  
NGC~2419 stands out from both the other GCs and the dwarf spheroidals. 
While \cite{coh12} suggested  that  being the nucleus of a disrupted dwarf 
galaxy could represent the explanation for the chemical signature of this 
cluster, no (present-day) dwarf presents Mg depletions as large as those 
observed in NGC~2419. 
On the other hand, such extremely Mg-poor stars are not observed even in 
the most massive GC in the Galaxy, $\omega$~Cen, whose Mg depletions are 
consistent with the pattern shown by dwarf spheroidals. 
Among the other GCs, only NGC~2808 shows three stars with very large Mg
depletions (at [Ca/H]$\sim -0.8$ dex, see Fig.~\ref{f:fig8}). 
These giants are likely the progeny of the most He-rich 
main-sequence population. 
We note, however, that not even these stars have comparably extreme [Ca/Mg] 
ratios as those observed in NGC~2419. 
The only star that resembles the Mg-poor population in NGC~2419 
is one giant in M~54 \citep{car10b}, the second most massive cluster 
in the Milky Way, sitting in the nucleus of the disrupting dwarf 
galaxy Sagittarius. 
Unfortunately, no K abundance is available for stars in M~54; 
a dedicated survey in this cluster could reveal precious information.

In conclusion, more massive objects, as well as systems with similar
metallicities, and stellar aggregates with different Galactocentric distances
do not seem able to match the extreme chemistry shown by a sizable part of the
stellar population of NGC~2419. 
It is possible that the peculiar combination of low metallicity, large mass, 
and large distance from the main parent Galaxy could explain the 
observed signatures of this object.
More observations of K for large samples of stars, like those assembled as
calibrators in \cite{kir11}, in a large number of GCs are required
to properly explore the mass-metallicity-distance parameter space.
Until then, NGC~2419 continues to be unique among GCs and old stellar systems.

\acknowledgements
Funding is acknowledged from  PRIN INAF 2011 "Multiple populations in
globular clusters: their role in the Galaxy assembly" (PI E. Carretta),  and
PRIN MIUR 2010-2011 ``The Chemical and Dynamical Evolution of the Milky Way and
Local Group Galaxies''  (PI F. Matteucci), prot. 2010LY5N2T. 
We are grateful also for funding from the U.S. National Science 
Foundation (grant AST-1211585).
We thank A.  Mucciarelli for sending us the unpublished EWs for K in NGC~2419, 
and I. Roederer for the EW of giants in M~22.

\clearpage

\clearpage

\clearpage

\clearpage

\clearpage

\clearpage

\clearpage

\clearpage

\clearpage

\begin{deluxetable}{rclrllllrcccll}
%\tabletypesize{\scriptsize}
\tabletypesize{\tiny}
\tablecaption{Relevant data for stars in the program clusters}%\label{data}}
\tablehead{\colhead{star}&
\colhead{GC.}&
\colhead{evol.}&
\colhead{S/N}&
\colhead{Teff}&
\colhead{logg}&
\colhead{[Fe/H]}&
\colhead{vt}&
\colhead{EW7698}&
\colhead{[K/Fe]}&
\colhead{[Mg/Fe]}&
\colhead{[Al/Fe]}&
\colhead{ref.}&
\colhead{ref.}\nl
\colhead{}&
\colhead{}&
\colhead{stage}&
\colhead{}&
\colhead{K}&
\colhead{dex}&
\colhead{dex}&
\colhead{km~s$^{-1}$}&
\colhead{m\AA}&
\colhead{NLTE}&
\colhead{dex}&
\colhead{dex}&
\colhead{par.atm.}&
\colhead{abund.}\\
}
\startdata
       &           &     &     &      &     &        &     &      &       &       &         &       &	 \nl
   435 & NGC~0104& SGB &  40 & 5190 &3.84 &$-$0.63 &0.00 &118.2 &$-$0.07& +0.31 &   +0.19 &   (a) & (1) \nl
   456 & NGC~0104& SGB &  40 & 5142 &3.84 &$-$0.68 &0.50 &136.3 &  +0.13& +0.32 &   +0.08 &   (a) & (1) \nl
   433 & NGC~0104& SGB &  40 & 5106 &3.84 &$-$0.74 &1.05 &158.2 &  +0.28& +0.49 &   +0.33 &   (a) & (1) \nl
   478 & NGC~0104& SGB &  30 & 5118 &3.84 &$-$0.56 &0.00 &138.8 &  +0.10&	  & $-$0.05 &   (a) & (1) \nl
201600 & NGC~0104& SGB &  40 & 5160 &3.84 &$-$0.61 &0.70 &136.3 &  +0.06& +0.33 &   +0.29 &   (a) & (1) \nl
   429 & NGC~0104& SGB &  35 & 5081 &3.84 &$-$0.61 &1.04 &160.0 &  +0.20& +0.39 &   +0.34 &   (a) & (1) \nl
201075 & NGC~0104& SGB &  40 & 5165 &3.84 &$-$0.64 &0.70 &138.6 &  +0.11& +0.42 &   +0.16 &   (a) & (1) \nl
206415 & NGC~0104& SGB &  40 & 5112 &3.84 &$-$0.67 &1.05 &161.0 &  +0.26& +0.34 &   +0.32 &   (a) & (1) \nl
  1012 & NGC~0104& TO  &  30 & 5832 &4.05 &$-$0.65 &1.07 &119.6 &  +0.16& +0.45 &   +0.09 &   (a) & (1) \nl
  1081 & NGC~0104& TO  &  45 & 5832 &4.05 &$-$0.66 &1.07 &118.9 &  +0.14& +0.50 & $-$0.10 &   (a) & (1) \nl
   975 & NGC~0104& TO  &  40 & 5832 &4.05 &$-$0.62 &1.07 &129.9 &  +0.27&	  & $-$0.19 &   (a) & (1) \nl
   482 & NGC~0104& SGB &  45 & 5090 &3.84 &$-$0.62 &0.84 &139.8 &$-$0.01& +0.64 &   +0.42 &   (a) & (1) \nl
     &           &      &   &    &    &      &  &      &      &      &       &       &        \nl
     113 &NGC~1904 &RGB & 154 & 4430 &1.34 &$-$1.54 &1.66 &137.0 &$-$0.15& +0.16 &   +1.05 &   (b) & (2) \nl 
     185 &NGC~1904 &RGB & 119 & 4596 &1.63 &$-$1.55 &1.73 &127.0 &$-$0.12& +0.36 &   +0.68 &   (b) & (2) \nl 

\enddata
\tablenotetext{a}{Table 1 is published in its entirety in the electronic 
edition of ApJL. A portion is shown here for guidance regarding its form and 
content.}
\tablenotetext{b}{Evolutionary stage: \nl
TO=turnoff; SGB=subgiant; RGB=red giant branch.}
\tablenotetext{c}{References for atmospheric parameters:\nl 
(a)=Carretta et al. (2004)\nl 
(b)=Carretta et al. (2009)\nl
(c)=Pancino et al. (2002)\nl
(d)=Gratton et al. (2001)\nl
(e)=unpublished/this work.}
\tablenotetext{d}{References for abundances:\nl 
(1)=Carretta et al. (2004: Mg, Al)\nl
(2)=Carretta et al. (2009: Mg, Al)\nl
(3)=Pancino et al. (2002: Fe)\nl 
(4)=this work (Mg, Al)\nl
(5)=Gratton et al. (2001)\nl
(6)=Carretta et al. (2005)}
\end{deluxetable}

\clearpage

\begin{deluxetable}{lllllrrl}
%\tabletypesize{\scriptsize}
\tabletypesize{\tiny}
\tablecaption{Relevant data for field stars}
\tablehead{\colhead{star}&
\colhead{Teff}&
\colhead{logg}&
\colhead{[Fe/H]}&
\colhead{vt}&
\colhead{EW7698}&
\colhead{[K/Fe]}&
\colhead{evol.}\nl
\colhead{}&
\colhead{K}&
\colhead{dex}&
\colhead{dex}&
\colhead{km~s$^{-1}$}&
\colhead{m\AA}&
\colhead{NLTE} &
\colhead{stage}\nl
}
\startdata
            &      &      &         &     &        &      &            \nl
HD 2665     & 5061 & 2.35 & $-$1.95 & 1.50&   37.7 & -0.57&   RGB-bump \nl   
HD 19445    & 6047 & 4.51 & $-$1.97 & 0.80&   21.0 & -0.09& 	  MS \nl   
HD 23439    & 5140 & 4.48 & $-$0.99 & 0.50&  155.0 &  0.34& 	  MS \nl   
HD 25329    & 4840 & 4.85 & $-$1.69 & 0.50&  121.0 &  0.20& 	  MS \nl   
HD 45282    & 5344 & 2.99 & $-$1.37 & 1.20&   91.0 &  0.04&  lower-RGB \nl   
HD 64090    & 5530 & 4.70 & $-$1.51 & 0.20&   59.0 & -0.09& 	  MS \nl   
HD 74462    & 4658 & 1.56 & $-$1.37 & 1.40&  144.0 &  0.20&  upper-RGB \nl   
HD 83212    & 4533 & 1.45 & $-$1.41 & 1.80&  161.0 &  0.07&  upper-RGB \nl   
HD 87140    & 5157 & 3.01 & $-$1.73 & 1.30&   87.0 &  0.15&  lower-RGB \nl   
HD 94028    & 6061 & 4.34 & $-$1.40 & 1.10&   64.0 &  0.06& 	  MS \nl   
HD 103095   & 5152 & 4.77 & $-$1.18 & 0.50&  121.5 &  0.08& 	  MS \nl   
HD 105546   & 5147 & 2.45 & $-$1.47 & 1.60&  113.0 &  0.15& 	 RHB \nl   
HD 110885   & 5382 & 2.54 & $-$1.45 & 1.90&  107.0 &  0.05& 	 RHB \nl   
HD 111721   & 4995 & 2.52 & $-$1.27 & 1.30&  122.0 &  0.16&  lower-RGB \nl   
HD 122956   & 4670 & 1.63 & $-$1.64 & 1.65&  120.8 & -0.02&  upper-RGB \nl   
HD 141531   & 4335 & 1.11 & $-$1.63 & 1.50&  152.0 &  0.09&  upper-RGB \nl   
HD 194598   & 6046 & 4.31 & $-$1.05 & 0.80&   81.0 &  0.12& 	  MS \nl   
HD 201891   & 5991 & 4.30 & $-$1.04 & 1.10&  101.0 &  0.30& 	  MS \nl   
LTT 10733   & 5757 & 4.37 & $-$0.92 & 1.10&  115.0 &  0.23&  lower-RGB \nl   
LTT 11819   & 5399 & 3.33 & $-$2.00 & 1.00&   42.0 & -0.12&  lower-RGB \nl   
BD +23 3912 & 5845 & 3.91 & $-$1.40 & 1.20&   76.5 &  0.13& 	  MS \nl   

\enddata
\tablenotetext{a}{Atmospheric parameters from \cite{gra00}.}
\end{deluxetable}

\clearpage

\begin{deluxetable}{lllllllllllll}
\tabletypesize{\scriptsize}
\tablecaption{Stars in common with Takeda et al. (2002)}
\tablehead{\colhead{star (HD)}&
\colhead{Teff}&
\colhead{logg}&
\colhead{[Fe/H]}&
\colhead{vt}&
\colhead{EW7698}&
\colhead{[K/Fe]}&
\colhead{Teff}&
\colhead{logg}&
\colhead{[Fe/H]}&
\colhead{vt}&
\colhead{EW7698}&
\colhead{[K/Fe]}\nl
}
\startdata
       &      &      &         &      &       &         &      &      &         &      &       &         \nl
       & \multicolumn{6}{c}{Takeda et al.}              &  \multicolumn{6}{c}{This work}                 \nl
103095 & 5000 & 4.50 & $-$1.15 & 1.00 & 118.0 & $-$0.15 & 5152 & 4.77 & $-$1.18 & 0.50 & 121.5 &   +0.08 \nl
122956 & 4660 & 1.75 & $-$1.93 & 1.70 & 120.8 &   +0.57 & 4670 & 1.63 & $-$1.64 & 1.65 & 120.8 & $-$0.02 \nl
201891 & 5827 & 4.43 & $-$1.04 & 1.60 & 101.4 &   +0.23 & 5991 & 4.30 & $-$1.04 & 1.10 & 101.0 &   +0.30 \nl
       &      &      &         &      &       &         &      &      &         &      &       &         \nl
\enddata
\end{deluxetable}

\end{document}